\documentclass[aps,prb,twocolumn,showpacs]{revtex4}
\usepackage[dvips]{graphicx}
\newcommand{\ifig}[2][]{\includegraphics[#1]{#2.eps}}
\usepackage{amssymb,amsmath,amsfonts}
\usepackage{wasysym}
\usepackage{latexsym}

\newcommand{\ket}[1]{| #1 \rangle}
\newcommand{\bra}[1]{\langle #1 |}

\newcommand{\be}{\begin{equation}}
\newcommand{\ee}{\end{equation}}

\newcommand{\qdot}{Quantum dot}
\begin{document}

\title{Semiclassical dynamics and long time asymptotics of the central-spin 
  problem in a quantum dot} 
\date{\today}

\author{Gang Chen$^1$, Doron L. Bergman$^1$, and Leon Balents$^1$}
\affiliation{${}^1$Department of Physics, University of California,
  Santa Barbara, CA 
  93106-9530}

\begin{abstract}
  The spin of an electron trapped in a quantum dot is a promising
  candidate implementation of a qubit for quantum information
  processing.  We study the central spin problem of the effect of the
  hyperfine interaction between such an electron and a large number of
  nuclear moments.  Using a spin coherent path integral, we show that in
  this limit the electron spin evolution is well described by
  classical dynamics of both the nuclear and electron spins.  We then
  introduce approximate yet systematic methods to analyze aspects of the
  classical dynamics, and discuss the importance of the exact
  integrability of the central spin Hamiltonian.  This is compared with
  numerical simulation.  Finally, we obtain the asymptotic long time
  decay of the electron spin polarization.  We show that this is
  insensitive to integrability, and determined instead by the transfer
  of angular momentum to very weakly coupled spins far from the center
  of the quantum dot.  The specific form of the decay is shown to depend
  sensitively on the form of the electronic wavefunction.
\end{abstract}
\date{\today}
\maketitle
  
\section{Introduction}\label{intro}
A single electron in a quantum dot provides a versatile experimental
realization of a single quantum bit (``qubit'').  The electron can be
trapped with a high degree of certainty by utilizing the Coulomb
blockade effect, in which a large charging energy prevents the motion of
electrons on or off the dot.  The two level degree of freedom of the
qubit is provided then by the electron spin.  The electron can be
readily manipulated electrically and optically.  A limiting factor in
such manipulations is ``decoherence'', caused by coupling of the
electron spin to other degrees of freedom.  In high-quality GaAs quantum
dots, in which all extrinsic sources of decoherence can be removed, the
hyperfine interaction with nuclear moments becomes
dominant.\cite{merkulov-2002-65,Khaetskii:prl02,Khaetskii2003} 

In a typical quantum dot, the electron spin is coupled to a very large
number $N\sim 10^4-10^6$ of nuclear moments, with a range of hyperfine
exchange constants determined by the wavefunction of the confined
electron.  This varies smoothly from near zero at the fringes of the dot
to a maximum value near the dot center.  To a rather good approximation,
there is no direct interaction between different nuclear moments.  The
latter is present only due to dipolar coupling, but is very small due to
the tiny nuclear dipole moment.  When it is neglected (which is expected
to be a good approximation for a relatively long time $\sim 10^{-3}sec$), the
Hamiltonian $\mathcal{H}$ of the electron-nuclear system has the
``central spin'' form, in which each nuclear spin operator
$\vec{I}_i$ (of spin $I$) is coupled only to the {\sl central}
electron spin ($1/2$) operator $\vec{S}$:
\begin{equation} \label{eqn1}
{\mathcal H}= \sum_{i} a_i \vec{I}_i \cdot \vec{S}
\; .
\end{equation}
The exchange constants $a_i$ vary with the amplitude of the electron
wavefunction in the \qdot, at the positions $\vec{r}_i$ of the nuclei:
$a_i \sim |\psi(\vec{r}_i)|^2$.

The central spin model is therefore of considerable experimental
relevance.  It is also of particular theoretical interest.  Most
intriguingly, it is known that electron spin decoherence within this
model is highly non-conventional.  For instance, the relaxation of an
initially polarized electron spin is {\sl non-exponential}.  Indeed,
numerical simulations for some specific forms of the electron
wavefunction indicated an inverse-logarithmic remanent polarization at
long times: \cite{Zhang2006,Al-Hassanieh2006,Erlingsson2004} 
\begin{equation}
  \label{eq:1}
  \langle S^z(t)\rangle \sim \frac{A}{|\ln t|^\alpha},
\end{equation}
with the initial condition $S^z(0)=+1/2$ and random unpolarized initial
conditions for the nuclear spin state (see Sec.~\ref{sec:formulation}); these numerical
results suggested $\alpha\approx 1$.

One possible reason to expect unusual behavior in the central spin
problem is that it is {\sl integrable}.  In fact, the central spin
Hamiltonian is a member of a large family of models studied by Gaudin,
possessing an infinite number of commuting constants of the motion, both
classically and quantum mechanically.  The Gaudin models unify the
central spin problem with another problem of recent interest, the
reduced BCS Hamiltonian for a small disordered superconducting grain.
This integrability has attracted additional theoretical
attention,\cite{Yuzbashyan2005,Yuzbashyan2005a} but as is often the
case, it is quite difficult to extract real-time dynamical properties
despite the model's formal solubility.  In any case, it is not clear
from these studies whether the unconventional long-time dynamics should
be ascribed to integrability or has another origin.

Another controversial aspect of the central spin problem is the
significance of {\sl quantum} coherence to the observed behavior.  The
pioneering studies in
Refs. \onlinecite{Khaetskii:prl02,Khaetskii2003,Khaetskii2003a} insist upon a
full quantum treatment, but employ approximations valid only in a large
applied magnetic field or for nearly full nuclear polarization.  The
work of Ref.\onlinecite{Erlingsson2004} instead argued on physical grounds
for the validity of the classical
approximation, and integrated the classical equations of
motion for the spins numerically, with however no assumptions on the
magnetic field or polarization (the focus was largely on the low field
and polarization situation most relevant to experiment).
Refs.\onlinecite{Al-Hassanieh2006,Zhang2006,Hasanieh:05} argue for a hybrid
approach in which {\sl effective} classical equations of motion are
employed using renormalized parameters, finding good agreement with
fully quantum simulations on small systems of $15-20$ spins.

In this paper, we will attempt to clarify the above issues through a
combination of techniques.  We will first argue, in agreement with
Ref.\onlinecite{Erlingsson2004}, that the classical approximation for the
nuclear spins is indeed justified {\sl for measurements of the
  electron spin dynamics} in the physically relevant large $N$ limit.
We show this by a simple direct analysis of the exact quantum expression
(``Keldysh-like form'') for the time-dependent expectation value of the
electron spin, thereby obtaining the explicit connection between the
properly defined quantum observable and the classical dynamics.  The
regime of validity of this approximation is discussed, and open issues
highlighted.  

Next, having established the correctness of the classical limit, we
attack the problem of the anomalous long-time spin dynamics.  Despite
the simplification of the classical limit, the dynamics remains highly
non-linear and non-trivial, and still is not easily accessed using
integrability -- though some progress has been made in this way for the
BCS problem\cite{Yuzbashyan2005,Yuzbashyan2005a,Yuzbashyan:cm04}.  We
will argue that the observed long-time behavior is due not to
integrability but rather to the effects of very weak but non-zero
coupling of the electron to spins far from the center of the dot, in the
tails of the electron wavefunction.  As time proceeds, successively more
weakly coupled nuclear spins are able to influence the electron spin
dynamics, leading to a slow change in the latter's behavior.  To address
the effects of integrability, we employ a phase space approach inspired
by fundamentals of statistical mechanics.  In general, one expects (the
ergodicity assumption) a Hamiltonian system to equally explore over
sufficiently long times all of its phase space accessible by the
constraints of its conservation laws.  An important feature particular
to the central spin problem is the presence of very weakly
coupled spins in the tails of the electron wave function.  Even for very
long times, some such spins have not yet been able to evolve
appreciably, further restricting the phase space which has been accessed
up to this time.  We capture this physics through a crude but
effective approximation. For any given time $t$, we divide the nuclear
spins into two groups: those with sufficiently large couplings to the
electron spin to allow them to precess through an angle of $O(1)$, and
those with weaker couplings that have not.  We then neglect completely
the latter ``decoupled'' spins, and treat the former ``coupled'' spins
as an isolated subsystem, and assume that it {\sl has fully explored its
accessible phase space already in the time $t$}.  In this ``instantaneous
ergodicity'' approximation, the long time dynamics is due to the
continual growth of the ergodic subsystem by inclusion of additional
weakly-coupled spins as time progresses.  

To study the properties of the ergodic average of the coupled subsystem
at any given time, one must still deal with the effects of
integrability.  Specifically, since the central spin problem is
integrable for any number of nuclear spins, the accessible phase space
is especially highly constrained due to the infinite number of constants
of the motion. We contend, however, that the constraints of
integrability are secondary to the already-described physics of weakly
coupled spins.  The basis of this conclusion is a re-organization of the
integrals of motion into an ordered set in which successive integrals
manifestly have less influence upon the electron spin.  The first four
(and therefore most important) integrals of motion in this set are the
three components of the total angular momentum and the energy.  We find
that {\sl ignoring} all remaining integrals of motion, i.e. replacing
averages over times less than $t$ by a uniform phase space average in
the coupled system at time $t$ {\sl without these constraints}, already
reproduces the observed non-exponential long-time spin relaxation.  We
show how additional integrals of motion can be systematically included
in this analysis, and compare the results to numerical simulations.

For the specific forms of wavefunction employed in prior numerical
simulations, an inverse logarithmic relaxation is obtained for long
times, in agreement with those numerical results.  This form is,
however, {\sl not} universal, and depends in particular upon the manner
in which the electron's wavefunction approaches zero.  The result can be
understood rather simply as follows.  Because the
electron is coupled to many nuclei, its characteristic precession
frequency is much larger than the rate at which any of the individual
nuclei themselves evolve, as they are coupled only to the single
electron spin.  Under this condition, as pointed out in
Ref. \onlinecite{Erlingsson2004}, the time-averaged electron spin simply
follows the evolution of the net hyperfine field
\begin{equation}
  \label{eq:46}
  {\boldsymbol{\sf H}}_N = \sum_j a_j {\bf I}_j.
\end{equation}
Because the hyperfine couplings $a_j$ decay with the distance of the
spin ${\bf I}_j$ from the center of the dot, the above hyperfine field
can be crudely approximated by including only those $N_s$ spins whose 
couplings are approximately constant and equal to the
maximal $a_{\rm max}$:
\begin{equation}
  \label{eq:center}
  {\boldsymbol{\sf H}}_N \approx a_{\rm max} \sum_{j=1}^{N_s} {\bf I}_j.
\end{equation}
Thus the hyperfine field is approximately proportional to the total
angular momentum of these $N_s$ strongest-coupled spins.  The total
angular momentum of {\sl all} spins is however conserved, and therefore
cannot decay but can only be redistributed.  Because there is no
preference for how this angular momentum is spread amongst the system's
spins, it is natural to expect that over long times, the total angular
momentum of all $N(t)$ spins which are appreciably involved in the dynamics is
uniformly distributed amongst them, and we have simply
\begin{equation}
  \label{eq:47}
  \langle S^z(t) \rangle \sim \langle {\bf{\sf H}}^z_N\rangle \sim
  \frac{N_s}{N(t)}.
\end{equation}
The time-dependence of $N(t)$ is determined from the fact that, at a
large time $t$, the dynamical spins are those which have had time to
precess, i.e. for which $a_j S t \gtrsim 1$.  

We can in this way readily understand the logarithmic behavior of
Eq.\eqref{eq:1} for the wavefunctions studied in
Refs.\onlinecite{Erlingsson2004},\onlinecite{Al-Hassanieh2006}.  For
these Gaussian and exponential wavefunctions,
\begin{equation}
  \label{eq:48}
 a_j \sim  |\psi(\vec{r}_j)|^2 \sim e^{-\left(\frac{r_j}{R_s}\right)^\gamma},
\end{equation}
where $R_s$ gives the ``radius'' of the dot and determines the number of
strongly coupled nuclear spins, $N_s \sim R_s^d$, for a $d$-dimensional
dot.  The parameter $\gamma=1,2$ for exponential and Gaussian
wavefunctions, respectively.  The dynamical spins at time $t$ are then
those inside some radius $R(t)$ such that 
\begin{equation}
  \label{eq:49}
  \frac{R(t)}{R_s} \sim (\ln t)^{1/\gamma}.
\end{equation}
$ $From Eq.\eqref{eq:47}, we then obtain Eq.\eqref{eq:1} with
$\alpha=d/\gamma$.  For the two-dimensional quantum dots ($d=2$) studied
in Refs.\onlinecite{Erlingsson2004},\onlinecite{Al-Hassanieh2006}, this
gives $\alpha=2,1$ for the exponential and Gaussian wavefunctions,
respectively.  In Sec.\ref{sec:asympt-behav-overl} we obtain these and
some other results from the above-described analytic calculations.

The remaining parts of the paper are organized as follows. In
Sec.~\ref{semiclassical}, we use spin coherent path integral to derive
the semi-classical equation of motion for both electron spin and nuclear
spins in the large $N$(number of nuclear spins) limit.  In
Sec.~\ref{sec:quasi-ergodic}, we make a short-time approximation to the
classical equations of motion and then focus on the nuclear subsystem.
Having systematically argued the relative importance of the integrals of
motion, we derive the long time average of effective nuclear field by
only considering the most important integrals of motion.  The results
from the numerical simulation and approximate predictions given in
Sec.~\ref{sec:quasi-ergodic} are compared in Sec.~\ref{sec:simulation}.
Finally in Sec.~\ref{sec:asympt-behav-overl}, with the approximation
explained above, the asymptotic time dependence of $\langle S^z(t)
\rangle$ is obtained for gaussian, exponentially decaying and hard wall
confining boundary coupling profiles.  We conclude in
Sec.\ref{sec:discussion} with a summary and discussion of future
prospects for theoretical development and comparison with experiment.

\section{Model and Semiclassical limit}\label{semiclassical}

\subsection{Formulation}\label{sec:formulation}

We begin by formulating a specific dynamical problem described by the
central spin Hamiltonian in Eq.\eqref{eqn1}.  Specifically, we consider
a general initial state in which the electron spin is polarized up:
\begin{equation}
 \label{ensemble_average}
 \ket{\Psi(0)} =
 \sum_{m_1,\cdots,m_N}{C_{m_1,\cdots,m_N}{\ket{m_1,\cdots,m_N}}_I
   {\ket{\tfrac{1}{2}}}_S} 
 \;,
\end{equation}
where subindex $I$,$S$ mean nuclear subsystem and electron,
respectively. And $I_i^z = m_i$.  The complex coefficients
$C_{m_1,\cdots,m_N}$ will be taken as independent zero mean Gaussian
random variables for each set of $m_i$.  The distribution is then fully
specified by the variance
\begin{equation}
  \label{eq:2}
  \overline{C^*_{m_1,\cdots,m_N}
    C_{m^{'}_1,\cdots,m^{'}_N}} = p(m_1,\cdots,m_N) \delta_{m_1m_1^{'}}
  \cdots \delta_{m_N m_N^{'}} ,
\end{equation}
where $p(m_1,\cdots,m_N)$ is a normalized ``classical'' probability
distribution for the z-components of the nuclear spins.  We will work
with the general form
\begin{equation}
  \label{eq:3}
  p(m_1,\cdots,m_N) = \frac{1}{Z} e^{h \sum_i m_i},
\end{equation}
where $Z=(\sinh [(I+1/2)h]/\sinh(h/2))^N$ by normalization.  The
parameter $h$ specifies the degree of initial nuclear polarization along
the $+z$ axis, with $h=0$ being unpolarized and $h\rightarrow \infty$ for
full polarization.

The remanent electron spin polarization is then defined as
\begin{equation}
  \label{eq:4}
  C(t) \equiv \langle S^z(t)\rangle = \left\langle
    \Psi(0)|S^z(t)|\Psi(0)\right\rangle, 
\end{equation}
where as usual, in the Heisenberg picture,
\begin{equation}
  \label{eq:5}
  \vec{S}(t) = e^{i\mathcal{H}t}\vec{S}e^{-i\mathcal{H}t}.
\end{equation}
It has been observed numerically\cite{Al-Hassanieh2006} that
$\langle\vec{S}(t)\rangle$ defined in this way is essentially {\sl
  self-averaging}, i.e independent of the realization of the
coefficients in Eq.\eqref{ensemble_average}. In
Appendix.~\ref{sec:approof},  we demonstrate that
this is indeed the case for $N\gg 1$, by showing that the normalized
fluctuations become {\sl exponentially} small in this limit:  
\begin{equation}
  \label{eq:6}
  \frac{\overline{\langle S^z(t)\rangle^2}- \left(\overline{\langle
        S^z(t)\rangle}\right)^2}{\left(\overline{\langle
        S^z(t)\rangle}\right)^2} \lesssim \left(\frac{\tanh
      (\frac{1}{2}h)}{\tanh[(I + \frac{1}{2})h] }\right)^N, 
\end{equation}
since the quantity in the brackets is always less than unity.  Therefore
one may approximate
\begin{equation}
  \label{eq:7}
  C(t) \approx \overline{C(t)} = \overline{\langle S^z(t)\rangle}.
\end{equation}
The latter averaged quantity is considerably simpler to treat
theoretically. 

Using Eqs.~(\ref{ensemble_average},\ref{eq:2},\ref{eq:3}) one obtains therefore
\begin{equation}
  \label{eq:8}
  C(t) \approx \frac{1}{Z} \left\langle{\tfrac{1}{2}}\right| 
        Tr_I [e^{-\tilde{\mathcal H}/2} e^{i{\mathcal H}t} S^z e^{-i{\mathcal
            H}t} e^{-\tilde{\mathcal H}/2}] \left|\tfrac{1}{2}\right\rangle_S,
\end{equation}
with the Boltzmann-like factor
\begin{equation}
  \label{eq:9}
  \tilde{\mathcal H} = - h \sum_i I_i^z.
\end{equation}
The partial trace and its contents in Eq.\eqref{eq:8} are extremely
reminiscent of a conventional equilibrium correlation function.  The
only difference is the presence of the electron spin operator inside the
Hamiltonian $\mathcal{H}$.  

To bring out this connection, we introduce a partial Trotter
decomposition for the exponentials.  We make use of the spin coherent
states, with the general form for spin $J$:
\begin{equation}
\ket{\hat n} = {\mathcal R}(\theta,\phi) \ket{J,+J}.
\end{equation}
Here the ${\mathcal R}(\theta,\phi)$ a the rotation operator that
changes the quantization axis from ${\hat z}$ to the direction defined
by the angles $(\theta,\phi)$.  The
resolution identity in this basis is
\begin{equation}
  \label{eq:10}
\frac{2 J+1}{4\pi} \int  d\hat{n} \ket{\hat n} \bra{\hat n} = 1 ,
\end{equation}
where the integration is over the directions on the unit sphere.  Similarly
the trace is given by 
\begin{equation}
  \label{eq:13}
  {\rm Tr} {\cal O} = \frac{2 J+1}{4\pi} \int  d\hat{n} \bra{\hat n}
  {\cal O} \ket{\hat n} .
\end{equation}
We first
use these coherent states {\sl for the electron spin only} to decompose
the exponentials of $\pm i{\mathcal H}t$, and obtain the path-integral
expressions
\begin{eqnarray}
  \label{eq:11}
  e^{-i{\mathcal H}t}  & = & \nonumber \\
  && \hspace{-0.5in} \int [d{\hat n}_+(t)] e^{i {\mathcal
      S}_B[\hat{n}_+]}
  \hat{T}^+_t e^{-i S \int_0^t \! dt' \hat{n}_{ +}(t') \cdot \vec{H}}
  |\hat{n}_{ +}(t)\rangle \langle
  \hat{n}_{ +}(0)|, \nonumber \\
  e^{+i{\mathcal H}t} & = &  \nonumber \\
  && \hspace{-0.5in} \int [d{\hat n}_{ -}(t)] e^{-i {\mathcal
      S}_B[\hat{n}_{ -}]} 
  \hat{T}^-_t e^{-i S \int_t^0 \! dt' \hat{n}_{ -}(t') \cdot \vec{H}}
  |\hat{n}_{ -}(0)\rangle \langle \hat{n}_{ -}(t)|, \nonumber \\
  \vec{H} & = & \sum_i a_i \vec{I}_i, 
\end{eqnarray}
where $\hat{T}^+_t$ ( $\hat{T}^-_t$) is the time-ordering
(anti-time-ordering) operator.  Here and in the following, we will use
square brackets to denote a functional integration measure.  The overlap
between coherent states in the Trotterization induces as usual a Berry
phase, 
\begin{equation}
  \label{eq:16}
  {\mathcal S}_B[\hat{n}] = S\int_0^t \! dt'\, \dot\phi \cos\theta,
\end{equation}
where $\hat{n} = (\sin\theta\cos\phi,\sin\theta\sin\phi,\cos\theta)$ and
$S=1/2$.  
Because the
different nuclear spin operators in each term of the sum defining
$\vec{H}$ (Eq.\eqref{eq:11}) commute, the operator exponentials can be
factored. 
% It is advantageous to express the trace in Eq.\eqref{eq:8}
% using spin coherent states, Eq.\eqref{eq:13}, for the nuclei, and separate the two
% exponential factors using the resolution of the identity in
% Eq.\eqref{eq:10}, again for the nuclei.   
Using then Eq.\eqref{eq:11} in Eq.\eqref{eq:8}, and expressing the trace
over the nuclear subspaces using Eq.\eqref{eq:13}, one obtains
\be\label{eq:20}
\begin{split} &
C(t) =  \frac{1}{\tilde Z} 
\int \prod_i d\hat{n}_i \int [d\hat{n}_{ +}  d\hat{n}_{ -}] 
\\ &  
e^{i ({\mathcal S}_B[\hat{n}_{ +}] - {\mathcal S}_B[\hat{n}_{ -}])}  
\langle\hat{n}_{ -} (t)| S^z |\hat{n}_{+} (t)\rangle
\\ &
\prod_i \big\langle \hat{n}_i \big| e^{ h I_i^z/2} U^{
      -}_i(0,t)U^{ +}_i(0,t)  e^{ h I_i^z/2}  \big|\hat{n}_i\big\rangle 
\; ,
\end{split}
\ee
where $\tilde{Z}=Z (4\pi/(2I+1))^N$, $|\hat{n}_i\rangle$ is a coherent
state for the nuclear spin $\vec{I}_i$, and 
\begin{eqnarray}
  \label{eq:21}
  U^\pm_i(t_1,t_2) & = & T_t^\pm e^{\mp i a_i S \int_{t_1}^{t_2} dt\, \hat{n}_{ \pm}(t) \cdot
      \vec{I}_i}.
\end{eqnarray}

Inspecting Eq.\eqref{eq:20}, we note that the integrand of the
functional integrals over $\hat{n}_\pm(t)$ contains a product of $N$
factors -- expectation values in the state $|\hat{n}_i\rangle$ -- one
for each individual nuclear subspace.  Each such factor is a {\sl
  functional} of $\hat{n}_\pm(t)$, and is distinguished from the others
{\sl only} by its value of $a_i$ (and the state $\hat{n}_i\rangle$).
Since there are many ($O(N)$) such factors all with $a_i$ of the same
order, this product is sharply-peaked/rapidly-oscillating and renders
the path integral suitable for a saddle point expansion (formally it can
be rewritten in the usual form as the exponential of a sum of $N$
smoothly-varying terms).  We will see below that the Berry phase terms
are also rapidly oscillating.

The saddle
point equations are:
\be\label{eq:15}
\begin{split} &
\frac{\delta}{\delta \hat{n}_{ \pm}(t')}
\Bigg[  
\sum_i \ln \big\langle \hat{n}_i\big|  
e^{ h I_i^z/2}  U^{-}_i(0,t)U^{ +}_i(0,t)  e^{ h I_i^z/2}  
\big|\hat{n}_i\big\rangle
\\ & \pm i{\mathcal S}_B[\hat{n}_{ \pm}] 
\Bigg] = 0
\; .  
\end{split}
\ee
This saddle point equation actually leads to the classical equations of
motion.  To see this, we evaluate the functional derivatives in
Eq.\eqref{eq:15}.   The variation of the Berry phase
term is well known~\cite{Auerbach}:
\begin{equation}
  \label{eq:17}
  \frac{\delta {\mathcal S}_B}{\delta
    \hat{n}_{ \pm}} = S \, \hat{n}_{\pm}
  \times{\dot{\hat{n}}}_{ \pm} .
\end{equation}
The variation of the expectation value is more involved:  
\be\label{eq:12}
\begin{split} &
\frac{\delta}{\delta \hat{n}_{\pm}(t')} 
\ln \bra{\hat{n}_i} 
e^{ h I_i^z /2} 
U^{ -}_i(0,t)  U^{ +}_i(0,t) 
e^{ h I_i^z /2} 
\ket{\hat{n}_i}
\\
& = 
\mp i a_i S \langle \vec{I}_i\rangle^{\pm}_{tt'} \; , 
\\
&
\langle \vec{I}_i \rangle^+_{tt'} = \frac{
\bra{\hat{n}_i} 
e^{ h I_i^z/2} 
U^{-}_i(0,t)  U^{+}_i(t',t)
\vec{I}_i
U^{+}_i(0,t') 
e^{ h I_i^z /2}
\ket{\hat{n}_i}
}{
\bra{\hat{n}_i}
e^{ h I_i^z/2} 
U^{-}_i(0,t)  U^{+}_i(0,t)  
e^{ h I_i^z /2}  
\ket{\hat{n}_i}}
\\
&
\langle \vec{I}_i \rangle^-_{tt'} = \frac{
\bra{\hat{n}_i} 
e^{ h I_i^z/2} 
U^{-}_i(0,t)  
\vec{I}_i
U^{-}_i(t',t)
U^{+}_i(0,t') 
e^{ h I_i^z /2}
\ket{\hat{n}_i}
}{
\bra{\hat{n}_i}
e^{ h I_i^z/2} 
U^{-}_i(0,t)  U^{+}_i(0,t)  
e^{ h I_i^z /2}  
\ket{\hat{n}_i}}
\; .
\end{split}
\ee

Thus the saddle point equations become:
\begin{eqnarray}
  \label{eq:14}
 \hat{n}_{ \pm}(t')
  \times{\dot{\hat{n}}}_{ \pm}(t') & = & \sum_i a_i \langle \vec{I}_i\rangle^\pm_{tt'} .
\end{eqnarray}
These equations are solved by the restricted condition
$\hat{n}_{ +}(t')=\hat{n}_{ -}(t')
\equiv \hat{n}(t')$.  We then accordingly write $U^{
  +}_i(t_1,t_2)\equiv U_i(t_1,t_2)$ and one has $U^{
  -}_i(t_1,t_2)=[U_i(t_1,t_2)]^\dagger$.  One obtains
\begin{equation}
  \label{eq:22}
  \langle \vec{I}_i\rangle^\pm_{tt'} =\frac{ \big\langle \hat{n}_i\big| e^{ h I_i^z /2} [U_i(0,t')]^\dagger \vec{I}_i
      U_i(0,t') e^{ h I_i^z /2} \big|\hat{n}_i\big\rangle }{\big\langle \hat{n}_i\big|e^{ h I_i^z } \big|\hat{n}_i\big\rangle }\equiv {\bf I}_i(t').
\end{equation}
The ``classical'' nuclear spin variable defined in this way has the form
of the expectation value of a standard Heisenberg operator.  Thus it
obeys
\begin{equation}
  \label{eq:23}
  \partial_t {\bf I}_i = a_i {\bf S} \times {\bf I}_i,
\end{equation}
with
\begin{equation}
  \label{eq:24}
  {\bf S} = S \hat{n},
\end{equation}
and the initial condition
\be\label{eq:25}
\begin{split} &
 I^{x,y}_i(t=0) = I \frac{n_i^{x,y}}{\cosh{(h/2)}+\sinh{(h/2)}n_i^z}, \\ &  
 I^z_i(t=0) = I \frac{\cosh{(h/2)}n_i^z + \sinh{(h/2)}}{\cosh{(h/2)}+\sinh{(h/2)}n_i^z} \;,
\end{split}
\ee
and ${\bf I}_i^2(0) = I^2$, in the unpolarized case when $h=0$, 
the above expression reduces to a simple form ${\bf I}_i(0) = I \hat{n}_i$.
Combining the results of Eqs.\eqref{eq:14},\eqref{eq:22} and the
definition in Eq.\eqref{eq:24}, the saddle point equation itself reduces
to
\begin{eqnarray}
  \label{eq:27}
  \partial_t {\bf S} & = &  {\boldsymbol{\sf H}}_N \times {\bf S}, \\
  {\boldsymbol{\sf H}}_N & = & \sum_i a_i {\bf I}_i.
\end{eqnarray}
Because of the polarized initial electron spin state, one has
\begin{equation}
  \label{eq:28}
  {\bf S}(t=0) = S \hat{z}.
\end{equation}
Note that Eqs.~(\ref{eq:23},\ref{eq:25},\ref{eq:27},\ref{eq:28}) are
precisely the 
classical equations of motion for this system, with the specified
initial conditions.  The remanent polarization is obtained from these
classical variables by evaluating the path integrals over
$\hat{n}_\pm(t')$ at the saddle point, and performing the remaining
integrations over the $\hat{n}_i$ variables:
\begin{equation}
  \label{eq:29}
   C(t)  = \left[ S^z(t;\{\hat{n}_i\}) \right].
\end{equation}
Here $S^z(t;\{\hat{n}_i\})$ is the solution to the classical equations
of motion for the nuclear initial conditions specified in
Eq.\eqref{eq:25}, evaluated at time $t$, and the square brackets
indicate an average over the nuclear initial conditions:
\begin{equation}
  \label{eq:30}
  \left[ A(\{\hat{n}_i\})\right] = \frac{\int \prod_i d\hat{n}_i [\cosh{(h/2)}+\sinh{(h/2)}n_i^z]^{2I} A(\{\hat{n}_i\})}
          {\int \prod_i d\hat{n}_i [\cosh{(h/2)}+\sinh{(h/2)}n_i^z]^{2I} }.
\end{equation}
The normalization is determined by the requirement that $C(0)=1/2$.

\section{Classical analysis}
\label{sec:quasi-ergodic}

\subsection{Reduced nuclear equations}
\label{sec:reduc-nucl-equat}
 
In the previous section we showed the classical equations of motion are
a good approximation for the dynamics of the system. In this section we
proceed to analyze them. The classical equations of motion read
\begin{equation}
\begin{split}
\partial_t {\bf S} & = {\boldsymbol{\sf H}}_N \times {\bf S}
\\
\partial_t {\bf I}_j & = a_j {\bf S} \times {\bf I}_j
\; ,
\end{split}
\end{equation}
where ${\boldsymbol{\sf H}}_N = \sum_j a_j {\bf I}_j$ acts as an effective magnetic field, which the electron spin
precesses about.  Here $|{\bf S}|=S=1/2$ and $|{\bf I}_j|=I$. 

Since the magnitude of the effective field ${\sf H}_N \equiv
|{\boldsymbol{\sf H}}_N|$ is of order $\sim \sqrt{N}$, while
an individual nuclear spin couples with only the single electron spin,
the timescale of the dynamics for the nuclear spins is much longer than
the electron spin.~\cite{Erlingsson:prb04}  As a result, electron spin
precesses rapidly about an  axis which itself moves only much more
slowly with time:
\begin{equation}
  \label{eq:19}
  {\bf S}(t) = \langle {\bf S}\rangle_t + \sqrt{S^2-|\langle {\bf
      S}\rangle_t|^2} {\rm Re}\left[ ({\bf e}_1+i
    {\bf e}_2) e^{i {\sf H}_N t}\right],
\end{equation}
where ${\bf e}_1,{\bf e}_2$ are orthogonal unit vectors in the plane
perpendicular to the short-time average of the electron spin $\langle
{\bf S}\rangle_t$,
\begin{equation}
  \label{eq:18}
  \langle {\bf S}\rangle_t = S \cos \theta_0 {\bf\hat m},
\end{equation}
which itself is along the  axis ${\bf\hat m}$ of the effective nuclear field
\begin{equation}
  \label{eq:26}
  {\bf\hat m} = \frac{{\boldsymbol{\sf H}}_N}{{\sf H}_N}.
\end{equation}
The vectors ${\bf\hat m},{\bf e}_1,{\bf e}_2$ form an orthonormal frame, 
and vary on timescales much slower than the electron precession quasi-period
 $1/{\sf H}_N$. The parameter $\theta_0$ is the
initial angle between the electron spin polarization and
$\boldsymbol{\sf H}_N(t=0)$.  To study the dynamics of the nuclear spin, it is
sufficient (for large $N$) to replace the electron spin by its time
average, leading to reduced equations of motion
\begin{equation}
  \label{nucs_only}
  \partial_t {\bf I}_j = g_j {\bf\hat m} \times {\bf I}_j,  
\end{equation}
With $g_j = S \cos\theta_0\, a_j$.  Eqs.\eqref{nucs_only}
eliminate the electron spin completely, and form a closed description of
the nuclear dynamics.  They are equivalent to the Hamiltonian
\begin{equation}
{\mathcal H}_{\textrm{eff}} =S \cos({\theta}_0) {\sf H}_N
\; ,
\end{equation}
which involves only the nuclear spins. 

\subsection{Conserved quantities}
\label{sec:conserved-quantities}

$ $From the Hamiltonian in Eq.\eqref{nucs_only}, it is
obvious the magnitude of the effective field ${\sf H}_N$ is a
\emph{conserved} quantity.  By spin-rotational invariance, the total
spin angular momentum is conserved, which in the above time-averaged
electron approximation becomes conservation of just the total nuclear
spin ${\boldsymbol{\sf I}} = \sum_i {{\bf I}_i}$ (the electron's
contribution to the total spin of the system is negligible for large
$N$).  It is readily verified that $\bf{\sf I}$ is indeed
conserved directly from Eq.\eqref{nucs_only}.

In fact one can find a series of conserved quantities for the reduced
equations of motion, obtained in the same way from those for the
complete central spin problem\cite{Yuzbashyan:cm04}
\begin{equation}\label{conserved_quanta}
{\mathcal H}_q = \sum_{j \neq q} \frac{2 g_j g_q }{g_j-g_q} {\bf I}_q \cdot {\bf I}_j
\;,
\end{equation}
where $q = 1, \ldots ,N$ runs over all the nuclear spins, and we have
assumed that no two nuclear spins have the same coupling $g_j$ to the
electron spin (this is a simplifying but inessential assumption for our
purposes).  

It will be convenient to consider a different linear combination of
these constants of the motion, which are naturally also conserved:
\begin{equation}
\label{eq:32}
{\mathcal C}_n = \sum_q {g_q^{n} {\mathcal H}_q}. 
% = A_n + \frac{1}{2} \sum_{k=0}^{n-1} {\bf h}_k \cdot {\bf h}_{n-1-k}
\end{equation}
The ${\mathcal C}_n$ can be defined for any $n$, and any set containing $N$ members
of ${\mathcal C}_n$'s 
is completely equivalent to the original set of $\{{\mathcal H}_q\}$, because 
the matrix determinant of this linear tranformation from $\{{\mathcal H}_q\}$ to $\{{\mathcal C}_n \}$
is nonvanishing, as all the couplings are different. Also notice that the
original set of  ${\mathcal H}_q$ is linearly independent {\sl except} for one constraint 
\begin{equation}
  \label{eq:33}
  {\mathcal C}_0 = \sum_{q=1}^n {\mathcal H}_q = 0,
\end{equation}
which follows from Eq.\eqref{conserved_quanta}. The energy $S\cos{({\theta}_0)}{\sf H}_N$
and magnitude of the total spin are contained in this set via
\begin{eqnarray}
  \label{eq:34}
  {\mathcal C}_{1}  & = & - S^2 \cos^2{(\theta_0)} {\sf H}_N^2 + I^2 \sum_j{g_j^2}, \\
  {\mathcal C}_{-1} & = &   {\sf I}^2 - I^2 N,
\end{eqnarray}
where ${\sf I}=|\bf{\sf I}|$.

\subsection{Collective variables and steady-state relations}
\label{sec:coll-vari-steady}

It is useful to define a set of vector quantities, which serve as
collective variables:
\begin{equation}
  \label{eq:35}
  {\bf h}_n \equiv \sum_j g_j^{n} {\bf I}_j.
\end{equation}
And $ {\bf h}_1 = S \cos{(\theta_0)} \boldsymbol{\sf H}_N$, $ {\bf h}_0 = \boldsymbol{\sf I}$.
The values of ${\bf h}_n$ determine the constants of motion ${\mathcal
  C}_n$ (but not vice-versa):
\begin{eqnarray}
  \label{eq:31}
 & n>0 &,\ \  {\mathcal C}_n =  - \sum_{k=1}^{n} {\bf h}_k \cdot {\bf
    h}_{n+1-k} + n I^2 \sum_j g_j^{n+1}
    \\
 & n<0 &,\ \  {\mathcal C}_n =    \sum_{k=0}^{-n-1} {\bf h}_{-k} \cdot {\bf
    h}_{n+1+k} + n I^2 \sum_j g_j^{n+1}
    \;.
\end{eqnarray}
In addition, the ${\bf h}_n$'s obey simple dynamical
equations obtained by taking moments of the equations of motion,
Eq.\eqref{nucs_only}: 
\begin{equation}
\label{eq:37}
\partial_t {\bf h}_n = \hat{\bf m} \times {\bf h}_{n+1}
\; .
\end{equation}

Here, we have expressed the spin dynamics completely in terms of these
collective variables.  Any set of $N$ contiguous collective variables, 
$\{ {\bf h}_k,\cdots,{\bf h}_{k+N-1}\}$ for any $k$, is equivalent to
the original spin variables.  This is because Eq.\eqref{eq:35} defines a
linear transformation of the van der Monde form, whose determinant is
therefore non-vanishing provided none of the $g_j$ are equal, as
assumed.  Furthermore, the field ${\bf h}_{k+N}$ appearing in the equation
of motion (Eq.\eqref{eq:37}) for $\partial_t {\bf h}_{k+N-1}$ can be eliminated in
favor of this set since it is not linearly independent of them, leading
to closed equations of motion for the collective variables.

The point of view in which the ${\bf h}_n$ are taken as our dynamical
variables is useful for elucidating the relative importance of the
${\mathcal C}_n$ for positive $n$.  The constants of motion of course
prevent the collective variables from ergodically exploring all of phase
space.  We are interested primarily in the nuclear hyperfine field ${\bf
  h}_1$.  From Eq.\eqref{eq:31}, clearly all ${\mathcal C}_n$ for $n>0$
involve ${\bf h}_1$, and thereby restrict somewhat its evolution.
However, the invariants for increasing $n$ involve progressively more
terms which {\sl do not} include ${\bf h}_1$.  Thus we expect the
corresponding constraints to be less restrictive on the average nuclear
field.

Conversely, to understand the (in)significance of invariants ${\mathcal
  C}_n$ with {\sl negative} $n$ is more readily seen in the original
nuclear spin variables.  Specifically, when $n<0$, greater weight is
given in ${\mathcal C}_n$ to those nuclear spins with the {\sl weakest}
couplings to the electron (and hence to each other in the effective
equation of motion).  Therefore we expect these integrals of motion to play
successively weaker roles in the dynamics in general, and especially in
the dynamics of the electron.  Consequently, we focus on the ${\mathcal
  C}_n$ with small positive $n$.

Suppose we consider the infinite time limit for a finite system
$N<\infty$.  Let us consider the time average of some physical quantity over
this time, defined by
\begin{equation}
\label{eq:36}
\langle {\cal O} \rangle \equiv 
\lim_{\tau \rightarrow \infty} \frac{1}{\tau}\int_0^{\tau}
\! dt\, {\cal O}(t) \; .
\end{equation} 
In ergodic systems, this limit is well-defined for most quantities.  In
integrable systems, the existence of this limit may depend upon which
quantity is averaged.  Consider the time average of Eq.\eqref{eq:37} for
$n = -1$.  Since ${\bf h}_0 = \boldsymbol{\sf I}$, it is conserved and thus
time-independent, and can be brought outside the time average. Therefore, 
one has
\begin{equation}
\label{eq:38}
\langle \partial_t {\bf h}_{-1} \rangle  = - \boldsymbol{\sf I}\times
\langle \hat{\boldsymbol{m}} \rangle 
\; .
\end{equation}
If we now assume both $\langle \hat{\boldsymbol{m}} \rangle $ and
$\langle {\bf h}_{-1} \rangle$ are well-defined, then the left-hand-side
in Eq.\eqref{eq:38} {\sl vanishes}.
  Thus we conclude $\boldsymbol{\sf I}\times
\langle \hat{\boldsymbol{m}} \rangle  =0$, i.e. the time average of the
nuclear field is parallel to the total angular momentum:
\begin{equation}
\label{eq:kappa}
\langle {\bf\hat m} \rangle =  \kappa \frac{\boldsymbol{\sf I}}{\sf I}
\; .
\end{equation}
The proportionality constant must satisfy $|\kappa| \leq  1$ since ${\bf
  \hat m}$ is a unit vector.  If $|\kappa| \neq 1$, this indicates
persistent motion of the nuclear field in the asymptotic time evolution.
Our numerics actually show that $\kappa$ is a convergent quantity.

\subsection{Approximate treatment of time averages}
\label{sec:appr-treatm-time}

Despite the very large number of integrals of motion, a complete
analytic solution of this problem has eluded the community.  In this
section, we adopt a less ambitious but more pragmatic approach, and
attempt to calculate the long-time average of the nuclear field
asymptotically.  We do this by first making the quasi-ergodic assumption
that this time average is equivalent to a uniform average over the
accessible regions of phase space.  The latter average is analogous to a
microcanonical ensemble average in statistical mechanics, but with more
constraints provided not only by energy conservation but by the many integrals of motion. 
 Mathematically, using the integrals
${\mathcal C}_n$, one has
\begin{eqnarray}\label{asymptotic_polarization}
\langle {\bf\hat m} \rangle_p &  = & 
\frac{1}{\mathcal Z}_p \int {\mathcal D} \left[ {\bf I}_j \right] {\bf\hat
  m}\,  
\delta\big(\sum_j {\bf I}_j - \boldsymbol{\sf I}\big) \delta(|\sum_j a_j
{\bf I}_j|-{\sf H}_N) \nonumber \\
& & \times \prod_{n=2}^p \delta\left( 
{\mathcal C}_n - c_n
\right)
\; ,
\end{eqnarray}
where $c_n$ are the fixed values of the integrals ${\mathcal C}_n$,
${\mathcal D}[{\bf I}_j] = \prod_j d{\bf I}_j$, and we have introduced
the partition function
\begin{eqnarray}
\label{eq:39}
{\mathcal Z}_p & = & \int {\mathcal D} \left[ {\bf I}_j \right]   
\delta\big(\sum_j {\bf I}_j - \boldsymbol{\sf I}\big) \delta(|\sum_j a_j
{\bf I}_j|-{\sf H}_N) \nonumber \\
& & \times \prod_{n=2}^p \delta\left( 
{\mathcal C}_n - c_n
\right)
\end{eqnarray}

In addition to this quasi-ergodic assumption, we make a separate
approximation, which consists of truncating the $O(N)$ series of
constraints in Eqs.(\eqref{asymptotic_polarization},\eqref{eq:39}) to only
the first few.  The degree of approximation is determined by the
parameter $p$ above.  For $p=1$, we include only energy and angular momentum
conservation, while $p>1$ counts the number of additional conserved
quantities taken into account.  This approximation is motivated by the
fact that, as argued above, the ${\mathcal C}_n$ are expected to have
increasingly little influence on the system's dynamics for increasing
positive $n$ and negative $n$.

To calculate $\langle {\bf\hat m}\rangle_p$ for small $p$, it is
convenient to introduce the (unnormalized) distribution of the {\sl vector} ${\bf\hat
  m}$: 
\begin{eqnarray}
  \label{eq:40}
&&  P_p({\bf\hat m};\boldsymbol{\sf I},{\sf H}_N)  = \int {\mathcal D} \left[ {\bf I}_j \right]   
\delta\big(\sum_j {\bf I}_j - \boldsymbol{\sf I}\big)  \\ & &   \times\delta\big(\sum_j a_j
{\bf I}_j-{\sf H}_N {\bf\hat m}\big)
 \prod_{n=2}^p \delta\left( 
{\mathcal C}_n - c_n \right). \nonumber
\end{eqnarray}

By rotational invariance, this distribution is actually a function only
of the angle $\theta$ between ${\bf\hat m}$ and $\boldsymbol{\sf I}$ and
not the absolute directions of either vector.  Thus we may write
\begin{equation}
  \label{eq:41}
  P_p({\bf\hat m};\boldsymbol{\sf I},{\sf H}_N)  = P_p(\cos\theta;{\sf
    I},{\sf H}_N), 
\end{equation}
with ${\bf\hat m}\cdot \hat{\boldsymbol{\sf I}}=\cos\theta \equiv u$.  The desired
expectation value is then obtained by a simple spherical average as
\begin{equation}
  \label{eq:42}
  \langle {\bf\hat m} \rangle_p = \frac{\int_{-1}^{1} \! du
   u P_p(u)}{\int_{-1}^{1} \! du
    P_p(u) } \hat{\boldsymbol{\sf I}},
\end{equation}
where we have suppressed the dependence on ${\sf I},h_1$.

The distribution $P_p(\cos\theta)$ can be calculated for small $p$ by a
saddle point technique (for $N \gg 1$).  Details are given in
Appendix ~\ref{ap1}.  For $p=1$, one obtains the very
simple distribution
\begin{equation}\label{distribution1}
P_1(\cos(\theta)) = e^{\alpha \cos(\theta)}
\; ,
\end{equation}
with
\begin{equation}
  \label{eq:43}
  \alpha = {\frac{3{\bar{a}}}{ \overline{a^2}-{\bar{a}}^2 }}
\frac{{\sf H}_N {\sf I}}{N I^2},
\end{equation}
and we define the moments of exchange couplings
\begin{equation}
  \label{eq:abar}
  \overline{a^n} = \frac{1}{N} \sum_{j=1}^N a_j^n.
\end{equation}
The resulting average direction from Eq.\eqref{eq:42} is then
\begin{equation}
\label{eq:two}
\langle {\bf\hat m} \rangle_1 = 
\hat{\boldsymbol{\sf I}} \left(
\coth{\alpha} - \frac{1}{\alpha}
\right)
\; .
\end{equation}
For $\alpha \ll 1$, $\langle {\bf\hat m} \rangle_1$ has an
even simpler expression:
\begin{equation}\label{taylor}
\langle {\bf\hat m} \rangle_1 =\hat{\boldsymbol{\sf I}} \frac{\alpha}{3}
\end{equation}

The first improvement upon the above result is to include one additional
conserved quantity, which gives (see Appendix \ref{ap1})
\begin{equation}
\label{distribution2}
P_2(\cos\theta) = e^{\gamma \cos\theta + \beta \cos^2\theta }
\; ,
\end{equation}
where
\begin{eqnarray}
  \label{eq:45}
\gamma & = & \frac{6{\sf I} {\sf H}_N^2 (\overline{a^2} \cdot \overline{a^3}-\bar{a}\overline{a^4})
    + \frac{3{\sf I}K}{(S\cos{\theta_0})^3} (\bar{a} \overline{a^3}-(\overline{a^2})^2)}
     {2 N {\sf H}_N I^2[(\overline{a^2})^3 - 2 \overline{a^2}\bar{a}\overline{a^3} + (\overline{a^3})^2
  - (\overline{a^2}-(\bar{a})^2)\overline{a^4}]} \nonumber \\
\beta &  = &  
  \frac{3{\sf I}^2(\bar{a}\overline{a^3}-(\overline{a^2})^2)^2/[I^2(\overline{a^2} - {\bar{a}}^2)]}
  {2N[(\overline{a^2})^3 - 2 \overline{a^2}\bar{a}\overline{a^3} + (\overline{a^3})^2
  - (\overline{a^2}-(\bar{a})^2)\overline{a^4}]
  },
\end{eqnarray}
with $K \equiv 2 {\bf h}_2 \cdot {\bf h}_1 = -{\mathcal C}_2 + 2 I^2 \sum_j g_j^3$ 
The resulting average of the nuclear field direction becomes
\begin{equation}
\label{eq:three}
\langle {\bf\hat m} \rangle_2 = 
\hat{\boldsymbol{\sf I}}
\left[
\frac{2 e^{\frac{\gamma ^2}{4 \beta }+\beta } \sinh(\gamma )}
   {\sqrt{\pi } \sqrt{\beta }
   \left[
   F\left(\frac{\gamma+2 \beta }{2 \sqrt{\beta }}\right)
   - 
   F\left(\frac{\gamma -2 \beta }{2\sqrt{\beta }}\right)
   \right]}
- \frac{\gamma}{2 \beta}
\right]
\; ,
\end{equation}
where $F(z)$ is the imaginary error function defined by $F(0) = 0$
and $F'(z) = \frac{2}{\sqrt{\pi}} e^{z^2}$. 

In principle, we can take into account more and more conserved quantities. 
However, the calculation becomes more and more cumbersome. 

%The approximate distribution is helpful in doing an ensemble average for many 
%identical system in which the electron spin can be prepared in a given initial state but the effective magnetic
%fields differ in the inital values, which will be presented in the last section. 
In the next section, we compare the predictions of this approximate calculation with numerical simulation
of the classical equations of motion.

\section{Numerical Simulation}
\label{sec:simulation}
In this section we briefly review the numerical simulation of the classical
equations of motion for the nuclear spins alone. The equation we are using is
Eq.\eqref{nucs_only}. We use a finite difference
discretization of the equations of motion to advance the nuclear spins from
the initial conditions to any subsequent time.

In order to protect the unit magnitude of the nuclear spins in the finite 
difference equations, we use a rotation matrix about the effective field 
$\boldsymbol{\sf H}_N$ to advance the nuclear spin states from $\{ {\bf I}_i(t) \}$ to 
$\{ {\bf I}_i(t+\Delta t) \}$. Rather than simply advance to the next time step,
we use the Euler method - we take the values of ${\bf I}_i(t+\Delta t)$,
calculate $\frac{1}{2}(\boldsymbol{\sf H}_N(t)+\boldsymbol{\sf H}_N(t+\Delta t))$, and use this
new averaged effective field value to produce new values for 
${\bf I}_i(t+\Delta t)$. This procedure is repeated a number of times
until ${\bf I}_i(t+\Delta t)$ is converged. Mathematically, this above process 
can be expressed compactly as follows, start with
$$ \{ {\bf I}_i(t) \} \ \mbox{and} \ {\boldsymbol{\sf H}}^0_N(t) = \sum_i {a_i {\bf I}_i(t)},$$
then repeat
\be
\begin{split} &
{\bf I}_i^k(t + \Delta t)  =  {\mathcal R}({\hat{\boldsymbol{\sf H}}}^k_N(t), g_i \Delta t) {\bf I}_i(t), \\ &  
{\boldsymbol{\sf H}}^k_N(t + \Delta t)  =  \sum_i {a_i {\bf I}_i^k(t + \Delta t)}, \\ &
{\boldsymbol{\sf H}}^{k+1}_N(t)  =  \frac{1}{2} ({\boldsymbol{\sf H}}^0_N(t) + {\boldsymbol{\sf H}}^k_N(t + \Delta t)),
\end{split}
\ee
until
$$ |{\bf I}_i^{k+1}(t + \Delta t) - {\bf I}_i^k(t + \Delta t)|< \epsilon \;, $$
where, the operation ${\mathcal R}(\hat{\bf m}, \phi)$ is the rotation 
about direction $\hat{\bf m}$ by angle $\phi$,
the integer variable $k$ is the times repeated, and $\epsilon$ is used to measure the convergence.

To verify correctness, we checked for conservation of energy and total 
angular momentum $\boldsymbol{\sf I}$ for all the data series we generated.
In our simulations we found that these are conserved at an accuracy of $\Delta E/E, |\Delta {\boldsymbol{\sf I}}|/{\sf I} \sim 0.1\%-0.2\% $.
% which is 
%good up to $1/1000$ in the timestep we use.
Numerical simulation for a number of special solvable cases were also carried out
(``solvable'' means they can be solved analytically after the short-time average for
electron spin introduced in Sec.~\ref{sec:reduc-nucl-equat}).
These consist of the problem of two nuclear spins with different coupling constants and 
the problem of many nuclear spins with identical coupling constants. 
In both cases, we find agreement between the simulations and the analytic results within the numerical accuracy.

\begin{figure}
\begin{center}
\ifig[height=2.0in]{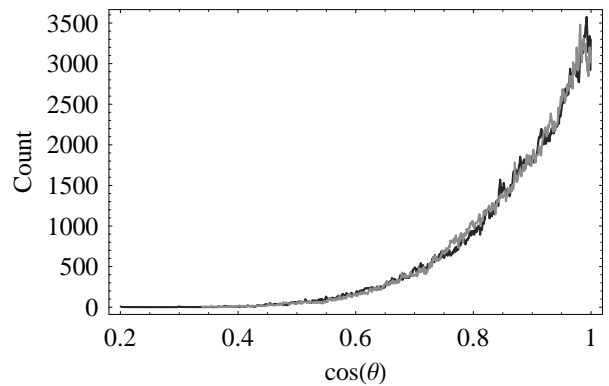}
\caption{Comparison between histogram profiles generated by $4th$ order Runge Kutta and our modified Euler method.
The dark gray profile is from data by $4th$ order Runge Kutta, the light gray one is from our modified Euler method. 
As expected, they almost coincide with each other.}
\label{fig:compare}
\end{center}
\end{figure}

We compare using the Euler method with a rotation matrix to the standard
finite difference $4th$ order Runge Kutta in our simulation by finding
the distribution $P(\cos{\theta})$ (introduced in Sec.\ref{sec:appr-treatm-time},
and measured by finding the histogram of the variable $\cos{\theta}$).
The two simulations were performed over $500,000$ timesteps of $0.05$, starting from the same initial conditions 
for $128$ nuclear spins with coupling constants uniformly distributed from $0$ to $1$,
The two measured distributions agree well, as plotted in Fig.\ref{fig:compare}, but the $4th$ order Runge Kutta
method requires much finer time discretization (smaller $\Delta t$) than our modified Euler method in order to preserve
energy and total angular momentum to the same degree of accuracy.
Furthermore, our modified Euler method more naturally conserves the
magnitude of each individual spin vector.

\begin{figure}
\begin{center}
\ifig[height=2.0in]{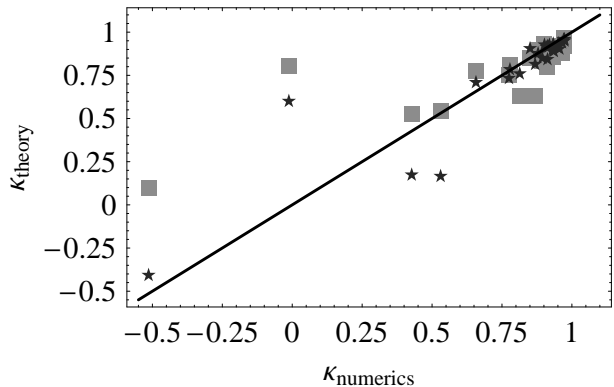}
\caption{Comparison of theory and numerics for values of $\kappa$.
The couplings are given by gaussian coupling profile $a_{i,j} = 2.0 \exp{[-\frac{\pi (i^2 + j^2)}{400}]}$,
where $i,j$ are integer values constrained by $i^2+j^2 \leq 900/{\pi}$, $900$ nuclear spins total.
The coordinate of each box(gray) is ($\kappa$-{numerics}, $\kappa$-{Eq.\eqref{eq:two}}); 
while the star(black) is ($\kappa$-{numerics}, $\kappa$-{Eq.\eqref{eq:three}}), same numerical value represents
same initial condition. The error bar of the numerical $\kappa$ is too small to be seen in this figure.
In this figure, $39$ different initial configurations are generated. 
The solid line is used as reference line, meaning the exact agreement between theory and numerics. 
(The same conventions hold for Fig.~\ref{fig:exponential},\ref{fig:gaussian2})}
\label{fig:gaussian1}
\end{center}
\end{figure}

\begin{figure}
\begin{center}
\ifig[height=2.0in]{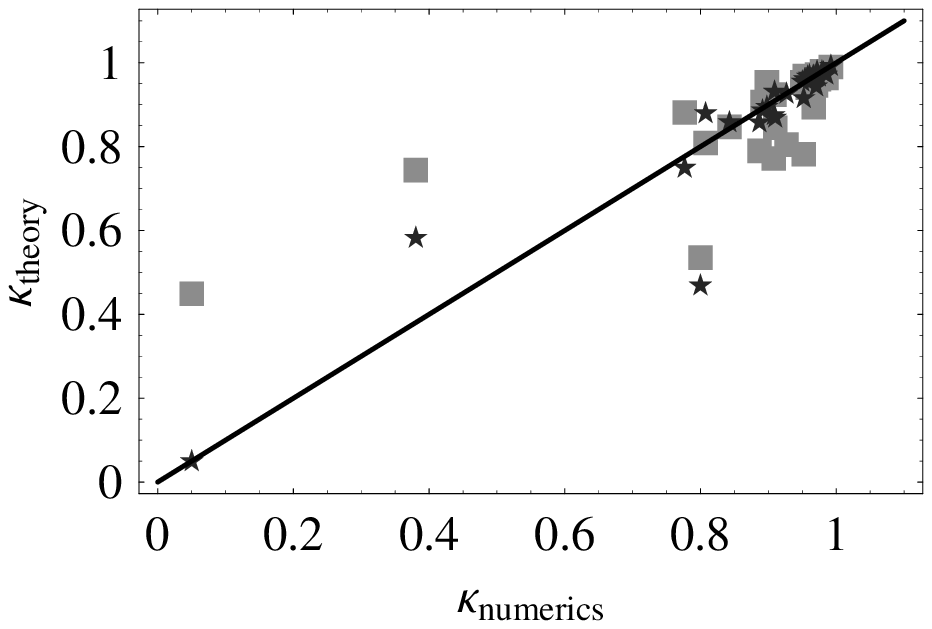}
\caption{Comparison of theory and numerics for values of $\kappa$.
The couplings are given by gaussian coupling profile $a_{i,j} = 2.0 \exp{[-\sqrt{\frac{\pi (i^2 + j^2)}{400}}]}$,
where $i,j$ are integer values constrained by $i^2+j^2 \leq 900/{\pi}$, $900$ nuclear spins total.
In this figure, $30$ different initial configurations are generated.
}
\label{fig:exponential}
\end{center}
\end{figure}

\begin{figure}
\begin{center}
\ifig[height=2.0in]{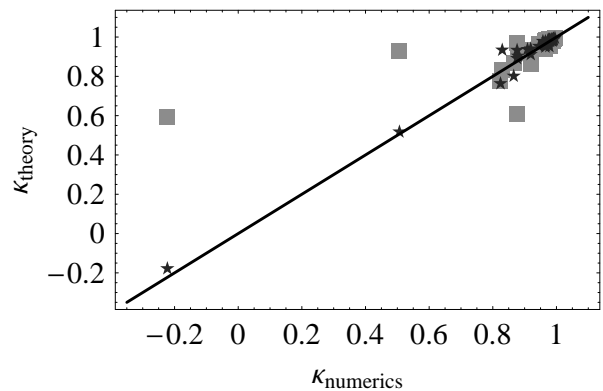}
\caption{Comparison of theory and numerics for values of $\kappa$.
The couplings are given by gaussian coupling profile $a_{i,j} = 2.0 \exp{[-\frac{\pi (i^2 + j^2)}{1000}]}$,
where $i,j$ are integer values constrained by $i^2+j^2 \leq 900/{\pi}$, $900$ nuclear spins total.
In this figure, $29$ different initial configurations are generated. 
}
\label{fig:gaussian2}
\end{center}
\end{figure}

In the simulation we find the complete time evolution of the variable $\cos{\theta}$.
We also find the value of $\kappa$ defined in Eq.\eqref{eq:kappa}.
The simulations are carried out for $600,000$ timesteps of $\Delta t = 0.1$, 
with different initial conditions and three different coupling profiles.
The approximate theoretical $\kappa$ values from Eq.\eqref{eq:two} and Eq.\eqref{eq:three} 
are plotted against the numerical $\kappa$ values in Fig.\ref{fig:gaussian1},\ref{fig:exponential},\ref{fig:gaussian2}
for the different coupling profiles. For each figure, the correlations between numerical data and two theoretical predictions
are explicitly calculated and listed in Table. ~\ref{tab:correlation}. For two data sets $X$ and $Y$, the correlation $\rho_{X,Y}$
between them is defined as
\begin{equation}
 \rho_{X,Y}=\frac{cov{(X,Y)}}{\sigma_X \sigma_Y}\;,
\end{equation}
where $cov{(X,Y)}$ is the covariance between $X$ and $Y$ and $\sigma_X$,$\sigma_Y$ are 
the standard deviation of data set $X$ and $Y$, respectively. The correlation is $1$ 
in the case of an increasing linear relationship, $-1$ in the case of a decreasing linear 
relationship, and some value in between in all other cases, indicating the degree of 
linear dependence between the variables. The closer the coefficient is to either $-1$
or $1$, the stronger the correlation between the variables. 
Although for some (very few) initial configurations, the analytical prediction given by Eq.\eqref{eq:three}
deviates from the numerical values more than Eq.\eqref{eq:two}, the correlations in the table clearly 
indicate that analytical prediction by Eq.\eqref{eq:three} is much better than Eq.\eqref{eq:two}. In addition,
Eq.\eqref{eq:two} always gives positive $\kappa$, Eq.\eqref{eq:three} can give negative $\kappa$ as shown by several
points in Fig.~\ref{fig:gaussian1},\ref{fig:gaussian2}, when the 
averaged effective nuclear magnetic field is anti-parallel with the total nuclear angular momentum $\bf{\sf I}$. 
The correlation table also indicates that,(though we need more numerics to confirm this indication) 
Eq.\eqref{eq:two} works better for the coupling profiles which are relatively 
more peaked, Eq.\eqref{eq:two} and Eq.\eqref{eq:three} works better for the coupling profiles which are more extended. 
Our explantion for this feature is that, the more peaked the coupling profile is, the better the approximation given by 
Eq.\eqref{eq:center} is; and under this approximation, the collective variables $\bf{h}_n$ defined by \eqref{eq:35} can 
be considered roughly proportional to $\boldsymbol{\sf H}_N$, so Eq.\eqref{eq:two} is probably enough to give relatively good prediction,
and Eq.\eqref{eq:three} won't improve the prediction significantly as for the more extended coupling profiles.

\begin{table}
 \begin{tabular}{|l|r|r|}
  \hline
  $\kappa$ data set            & numerics and Eq.\eqref{eq:two}  & numerics and Eq.\eqref{eq:three} \\
  \hline 
   Fig.\ref{fig:gaussian1}     & 0.817407  & 0.891817 \\
  \hline
   Fig.\ref{fig:exponential}   & 0.771773  & 0.926312 \\
  \hline
   Fig.\ref{fig:gaussian2}     & 0.704288  & 0.991488 \\
  \hline
 \end{tabular}
      \caption{Correlation between $\kappa$ value data from numerics and two theoretical 
      predictions given by Eq.\eqref{eq:two} and Eq.\eqref{eq:three} for three different coupling profiles,
      the data are the same as the ones used in the corresponding figures of the left column above.}
      \label{tab:correlation}
\end{table}

\section{Asymptotic behavior of $\langle S^z(t) \rangle$} 
\label{sec:asympt-behav-overl}

As discussed in the Introduction, even at long times $t$ there will be
some very weakly coupled spins which have not evolved appreciably.
These spins, being very weakly coupled to the electron, also do not
significantly contribute to its average or to the nuclear hyperfine
field.  Thus we need consider only the dynamics of those spins with
couplings $a_i t \geq k$ with some constant $k$ of $O(1)$.  We will keep
$k$ as a free parameter in the considerations below to assess the
sensitivity of our results to this choice.  Quantities which are
sensitive to $O(1)$ changes of $k$ are clearly not reliably obtained
from our approximations.

Let us for the time being focus on wavefunctions with the (stretched)
exponential form of Eq.\eqref{eq:48}, for which the size of the region
of dynamically coupled spins evolves logarithmically, as given in
Eq.\eqref{eq:49}.  For such logarithmic growth, the number of dynamical
spins increases extremely slowly at long times, with very few additional
spins being added with each proportional increase (e.g. doubling) in
time.  Thus we believe that the most of the spins within this region have
to a very good approximation explored their accessible phase
space -- subject of course to the constraints of integrability.  This is
the physical argument justifying the use of
Eq.\eqref{asymptotic_polarization} with the time-dependence only
appearing through the slow increase of the set of dynamical couplings.  

In Sec.~\ref{sec:coll-vari-steady}, we argued that invariants ${\mathcal
  C}_n$ have decreasing relevance to the electron dynamics with
increasing positive $n$.  In the more specialized context of this
section, we can slightly refine this contention.  Specifically, consider
the average $\langle \hat{\bf m}_p \rangle$ defined in
Eq.\eqref{asymptotic_polarization} in the long-time limit.  We are
interested only in the time-dependence of this quantity, i.e. how it
changes as additional very weakly coupled spins ${\bf I}_j$ are
introduced into the average.  These appear in the integration measure,
inside the definition of ${\bf\hat m}$, and in the $\delta$-function
constraints due to the constants of the motion.  However, in all these
places {\sl except} the measure and the $\delta$-function constraining
the total angular momentum $\boldsymbol{\sf I}$ -- and specifically in
all higher invariants ${\mathcal C}_n$ with $n\geq 2$ -- the additional
introduced spins are weighted by positive powers of their very small
couplings $a_j$.  Thus we expect that all these approximations give very
similar long-time dependence for $p\geq 1$.  Indeed, by explicit
computation (using e.g. Eq.\eqref{distribution2} and its
generalizations) we found identical results for the leading asymptotic
long-time behavior for $p=1,2$.

We therefore discuss the simplest but sufficient case $p=1$, in which
only energy and total angular momentum conservation are taken into
account.  To use Eq.\eqref{eq:two} we require $\alpha$ from
Eq.\eqref{eq:43}, which depends upon the first two moments of the
hyperfine couplings.  Consider the case of a Gaussian wavefunction,
$\gamma=2$ in Eq.\eqref{eq:48}.
Including those nuclear spins at radii $r <
R(t)$, one obtains ${\bar a} = \frac{a_0}{\ln{(t/k)}}$ and
${\overline{a^2}} = \frac{a_0^2}{2 \ln{(t/k)}}$, where $a_0$
gives the overall scale of the hyperfine coupling. 
A typical (root mean 
square ensemble average) value of the combination $\frac{{\sf H}_N {\sf I}}{N I^2} \sim
\sqrt{\overline {a^2}} \sim \frac{1}{\sqrt{\ln{(t/k)}}}$,
and for this reason the parameter $\alpha$
\begin{equation}
\alpha \sim \sqrt{\frac{1}{\ln{(\frac{t}{k})}}} \ll 1.
\end{equation}
Assuming $S^z(0)=1/2$, then $\cos{{\theta}_0} = \frac{{\sf H}_N^z}{{\sf
    H}_N} $ and we can evaluate the appropriate ensemble average derived
in Eq.\eqref{eq:29}, replacing the classical spin by its short-time
average and using Eq.\eqref{eq:18},\eqref{taylor}:
\begin{eqnarray}
 \label{longtime}
C(t) = \langle S^z(t)\rangle
                    &=& \left[\frac{1}{2} {\frac{{{\sf H}_N^z}(t)}{{{\sf H}_N}(t)} \frac{{\sf I}^z(t)}{{\sf I}(t)} \frac{\alpha}{3}}\right] \nonumber \\
                    &=& \left[\frac{1}{2} \frac{\bar{a}}{\overline{a^2}-\bar{a}^2} \frac{{{\sf H}_N^z}(t){\sf I}^z(t)}{NI^2}\right] \nonumber \\
                    &=& \frac{1}{6} \frac{\bar{a}^2}{\overline{a^2}-{\bar{a}}^2} \nonumber \\
                    &=& \frac{1}{6} \frac{1}{\ln{\frac{t}{k}}}
         \;,
\end{eqnarray}
where we have replaced ${{\sf H}_N^z}/{{\sf H}_N}$ with the effective
${{\sf H}_N^z(t)}/{{\sf H}_N(t)}$ found within $R(t)$ in the first line of the
above equations, as they are very close in the asymptotic time limit, and in
the fourth line, the ensemble averaging is taken, where $[{{\sf H}_N^z}(t){\sf I}^z(t)] = \bar{a} NI^2/3$ is used.
The asymptotic logarithmic decay agrees with the heuristic result of the introduction,
and with the numerical findings of Ref.~\onlinecite{Erlingsson:prb04,
Hasanieh:05}. Interestingly, the choice of $k$ {\sl does not} affect
the prefactor of the leading logarithmic behavior in
Eq.\eqref{longtime}, suggesting that not only the logarithmic dependence
but its prefactor might be properly obtained by this approach.  The
prefactor in \eqref{longtime} is indeed of the same order of magnitude as found
in the $1/S^z$-$\ln{t}$ plot by Ref.~\onlinecite{Hasanieh:05}, whose
treatment is equivalent to a classical one with $I = S = \sqrt{3}/2$ in
the equations of motion rather than $I = S = 1/2$.

We have carried out a similar calculation for an exponential
wavefunction, $\gamma=1$ in Eq.\eqref{eq:48}.  In this case, we obtain
\begin{eqnarray}
C(t)= \frac{4}{3} \frac{1}{{\ln^2{(\frac{t}{k})}}} 
\;,\label{eq:50}
\end{eqnarray}
which is different from the asymptotic decay for the previous example,
but agrees again with the arguments in the introduction. This disagrees
with the suggestion in Ref.\onlinecite{Hasanieh:05} that the decay in
this case would have the {\sl same} form as for the Gaussian
wavefunction above.  However, no justification was given in
Ref.\onlinecite{Hasanieh:05} for this claim apart from a slow growth of
$1/\langle S^z(t)\rangle$ on a plot against $\ln t$.  Given the
spread in their numerical data, it is likely that the form of
Eq.\eqref{eq:50} would produce a good fit.  

For the case of an infinitely high potential barrier in the quantum dot,
the coupling constants vanishes at the boundary, and the electron only
couples with a finite number of nuclear spin.  The nuclear spins near
the boundary have a coupling that vanishes quadratically with their
distance from the boundary $ g(r) \sim (1-r/l)^2$, $l$ being the radius
of the confining barrier.  Repeating the calculation above for this
case, we find
\begin{equation}
C(t) \sim c_1 + c_2 \sqrt{\frac{k}{t}}
\; ,
\end{equation}
where $c_1$ and $c_2$ are constants. This result can also be understood
from the argument given in the Introduction. When $R(t)$ reaches the
boundary of the confining potential, $R(t) = l(1-\sqrt{k/t})$ and $N(t)
= N (1-\sqrt{k/t})^2$, where $N$ is the total number of nuclear spins.
Plugging $N(t)$ into Eq.\eqref{eq:47}  leads to the above result.

\section{Discussion}
\label{sec:discussion}

In this article we have derived the appropriate classical limit for the
quantum central spin problem of a single electron spin coupled to many
nuclear spins, and shown that this limit is indeed valid when the number
of nuclei is large.  This validates the starting point of
Ref.~\onlinecite{Erlingsson:prb04}, which simply assumed this.
Furthermore, we have made some approximate analytic predictions for the
asymptotic behavior of the central (electron) spin in this system, and
have shown by numerical simulation that these predictions are roughly
correct.  Moreover, the approximations can be systematically improved,
as we also described, though the technicalities involved in this
improvement appear formidable.  In the long-time limit for a finite
number of nuclear spins, we argued that the asymptotic polarization of
the electron spin will on average point in the direction of the total
nuclear angular momentum $\boldsymbol{\sf I}$, regardless of the coupling constants
$a_j$.  Finally, we have analytically derived the functional form of the
decay of the electron spin polarization toward its asymptotic value.
This form depends on the coupling constant profile, and is in
qualitative and semi-quantitative agreement with observations from prior
numerical investigations.  Our results elucidates the physical origin of
the long-time remanent spin polarization, and indicates that,
remarkably, it is not connected to the integrability of the central spin
problem.  

The quite simple explanation of the long-time spin dynamics suggests
that numerous extensions of these results should be possible, which are
interesting in light of the relevance of the problem to experiments in
quantum dots.  Most obviously, one could likely explain the long-time
behavior in situatiuons with substantial nuclear polarization and/or
applied magnetic fields, such as treated in some special cases in
Refs.\onlinecite{Khaetskii2003,Khaetskii:prl02}.  Our derivation of the
classical limit in Sec.\ref{semiclassical} was general enough to apply
to these cases, so a treatment of this problem may proceed already from
this point.  Since integrability appears not to play a significant role,
there also appears to be no substantial obstacle to analysis of more
complex (and non-integrable) situations such as two electrons in near
proximity coupled to overlapping nuclear populations, and crossover
effects induced by weak nuclear dipole interactions.  It would also
obviously be interesting to treat more complicated dynamical quantities
than the remanent electron spin described here, such as higher order
spin autocorrelations, electron-nuclear entanglement, and the influence
of time-dependent fields.

The more important and probably more formidable challenge is to connect
the intriguing dynamical behavior of such electron-nuclear interaction
problems to experiment.  As was mentioned in the Introduction, the
hyperfine interaction is the dominant source of dephasing in current
single-spin experiments in GaAs quantum dots.  A recent review can be
found in Ref.\onlinecite{hanson-2006}.  Existing experiments use
optical\cite{Braun2005a,Dutt2005} or
electrical\cite{Johnson2005,Koppens2005} means to measure this
dephasing.  These experiments, however, focus upon the short-time
behavior, on the order of the electron precession frequency, in which
the nuclear field can be regarded as approximately static.  The
non-trivial behavior which is the focus of this paper appears only on
longer timescales.  It is possible to use spin-echo techniques to remove the
trivial dephasing effect of this fast electron precession about the
(uncontrolled) quasi-static nuclear
field\cite{Petta2005,Greilich2006b}.  This technique (and possibly
others) should in principle allow the unimpeded observation of
the slower nuclear dynamics.  Very recently, optical measurements have
achieved the sensitivity necessary to measure a single electron spin in
a {\sl single} quantum dot.\cite{Berezovsky2006,Atature2007}   Direct
observation of the electron spin in real time can of course also detect
both short and long time dynamics.  Given the
sometimes frantic pace of experimental developments in this field, we
hope that experimentalists as well as theorists will make the effort to
explore some of this very interesting physics before rushing to more
complex structures.  It is conceivable that the confirmation of our
physical picture of electron-nuclear spin interactions could even lead
to useful ideas for future devices.  For instance, the observation that
the coupling profile determines the form of the electron spin
polarization decay could enable control of time-dependent spin
polarization by quantum dot shape engineering.

\section{Acknowledgment}

We acknowledge discussions with V. Dobrovitski, and thank L. Glazman and
D. Loss for stimulating our interest in the problem.  This work was
supported by the Packard Foundation and the National Science Foundation
through grant DMR04-57440.

\appendix

\section{Self-Averaging of electron spin}
\label{sec:approof}
In this appendix, we show that the fluctuations of the dynamics in initial conditions
of the nuclear spin bath are negligible, thus demonstrating that the
system is self-averaging. We will do so by proving Eq.\eqref{eq:6}. 

We first consider $h=0$ in Eq.\eqref{eq:3}, in which case 
\begin{eqnarray}
\label{eq:a1}
\overline{\langle S^z(t)\rangle^2}  =  \sum_{\{m_i,m^{'}_i,n_i,n^{'}_i\}}
\overline{ C^{*}_{\{m_i\}} C^{*}_{\{n_i\}} C_{\{m^{'}_i\}}C_{\{n^{'}_i\}} } \nonumber \\
  \bra{\{m_i\}}\bra{\tfrac{1}{2}} S^z(t)\ket{\tfrac{1}{2}}\ket{\{m_i^{'} \}}
 \bra{\{n_i\}}\bra{\tfrac{1}{2}} S^z(t)\ket{\tfrac{1}{2}}\ket{\{n_i^{'} \}}
 \;,
 \end{eqnarray}
where $Z = (2I+1)^N $,
and
\begin{equation}
\label{eq:a2}
\overline{ C^{*}_{\{m_i\}} C^{*}_{\{n_i\}} C_{\{m^{'}_i\}}C_{\{n^{'}_i\}} } =
 \frac{\prod_i{\delta_{m_i,m_i^{'}}\delta_{n_i,n_i^{'}}}+\prod_i{\delta_{m_i,n_i^{'}}\delta_{m_i^{'},n_i}}}{Z^2},
\end{equation}

With Eqs. \eqref{eq:a1}, \eqref{eq:a2}, we can express the variance of the spin expectation value,
$\overline{\langle S^z(t)\rangle^2} - {\overline{\langle S^z(t)\rangle}}^2 $ as
\be
\label{eq:a3}
\frac{1}{Z^2}
\sum_{\{m_i\}} \bra{\{m_i\}}\bra{\tfrac{1}{2}} S^z(t)\ket{\tfrac{1}{2}} \bra{\tfrac{1}{2}} S^z(t)\ket{\tfrac{1}{2}}\ket{\{m_i\}}
\ee

In Eq.\eqref{eq:a3}, we have $\bra{\tfrac{1}{2}} S^z(t)\ket{\tfrac{1}{2}}^2 \leq \frac{1}{4}$, 
resulting in a variance of
\be
\overline{\langle S^z(t)\rangle^2} - {\overline{\langle S^z(t)\rangle}}^2 \leq 
\frac{1}{4 Z^2}
\sum_{\{m_i\}} \bra{\{m_i\}} \{m_i\} \rangle
=
\frac{1}{4 Z}
\ee
so the variance is of order $\sim Z^{-1} = (2I+1)^{-N}$. Since ${\overline{\langle S^z(t)\rangle}}^2 \sim O(1)$, we
find the $h \rightarrow 0$ limit of Eq.\eqref{eq:6}.

For the $h \neq 0$ case, an analogous derivation may be carried out, now including the Boltzmann-like factor
$\tilde{\mathcal H} = - h \sum_i I_i^z$ as defined by Eq.\eqref{eq:9}. The variance now becomes
\be
\begin{split} &
\overline{\langle S^z(t)\rangle^2} - {\overline{\langle S^z(t)\rangle}}^2 
\\ & =  
\frac{1}{Z^2}
\sum_{\{m_i,n_i\}}
| \bra{ \{m_i\} , \tfrac{1}{2} } 
S^z(t)
\ket{ \tfrac{1}{2} , \{n_i\} } |^2
e^{h \sum_i m_i + h \sum_i n_i} 
\\ & = 
\frac{1}{Z^2} 
\sum_{\{m_i\}}
\bra{\{m_i\}}
e^{-\tilde{\mathcal H}}
\bra{\tfrac{1}{2}}
  S^z(t)  \ket{\tfrac{1}{2}}
e^{-\tilde{\mathcal H}}
\bra{\tfrac{1}{2}}  S^z(t)  \ket{\tfrac{1}{2}}
\ket{\{m_i\}}
 \; .
\end{split}
\ee
Using $\bra{\tfrac{1}{2}}  S^z(t)  \ket{\tfrac{1}{2}} \leq \frac{1}{2}$
we find a bound on the variance
\be
\begin{split} &
\overline{\langle S^z(t)\rangle^2} - {\overline{\langle S^z(t)\rangle}}^2 
\\ & \leq  
\frac{1}{4 Z^2} 
\sum_{\{m_i\}}
\bra{\{m_i\}}
e^{-2 \tilde{\mathcal H}}
\ket{\{m_i\}}
\\ & = 
\frac{1}{4 Z^2}  \left( \frac{\sinh((2I+1)h)}{\sinh(h)} \right)^N
\\ & = 
\frac{1}{4}  \left( \frac{\tanh(\frac{h}{2})}{\tanh((I+\frac{1}{2})h)} \right)^N
 \; .
\end{split}
\ee
With ${\overline{\langle S^z(t)\rangle}}^2 \sim O(1)$, we
have Eq.\eqref{eq:6}. 

Since $\frac{\tanh{(\frac{1}{2}h)}}{\tanh{((I+\frac{1}{2})h)}} < 1$, in the large $N$ limit, 
the above expression decays to zero exponentially.

\section{Approximate distribution function for the effective nuclear field direction}
\label{ap1}
In this appendix we find the distribution function for calculating
\eqref{asymptotic_polarization}, taking into account only the two most significant 
conserved quantities.
\begin{equation}
{\mathcal Z}_1 = \int d{ \hat{\bf m} } 
\int {\mathcal D} \left[ {\bf I}_j \right]
\delta({\boldsymbol{\sf H}_N} - \sum_i a_i {\bf I}_i)\delta(\boldsymbol{\sf I} - \sum_i {\bf I}_i)
\; ,
\end{equation}
where $\hat{m}$ is the directional vector of $\boldsymbol{\sf H}_N$.
First we use Lagrange multipliers to get the form
\begin{equation}
{\mathcal Z}_1 = \int d{\hat{\bf m}} 
\int d{\bf x} d{\bf y} 
e^{i {\bf x} \cdot {\boldsymbol{\sf H}_N} + i {\bf y} \cdot \boldsymbol{\sf I} } 
\int {\mathcal D} \left[ {\bf I}_j \right] 
e^{-i \sum_i ({\bf x} \cdot  a_i {\bf I}_i + {\bf y} \cdot {\bf I}_i)}
\; .
\end{equation}
Next we integrate over the ${\bf I}_j$ variables
\begin{equation}
{\mathcal Z}_1 = \int d{\hat{\bf m}} 
\int d{\bf x} d{\bf y} 
e^{i {\bf x} \cdot {\boldsymbol{\sf H}_N} + i {\bf y} \cdot \boldsymbol{\sf I} } 
\prod_i
\frac{\sin(I| a_i {\bf x} + {\bf y} |)}{I| a_i {\bf x} + {\bf y} |}
\; ,
\end{equation}
where $I$ is the magnitude of the nuclear spin vector.
Now we use the thermodynamically large number of nuclear spins to first
approximate the sum over nuclear spins by a distribution function $p(a)$ for 
the coupling constants, and then to justify a saddle point approximation
\begin{equation}
\begin{split} &
\prod_i
\frac{\sin(I | a_i {\bf x} + {\bf y} |)}{I| a_i {\bf x} + {\bf y} |}
=
e^{N \int da p(a) \ln \left( 
\frac{\sin(I| a {\bf x} + {\bf y} |)}{I | a {\bf x} + {\bf y} |}
\right)}
\\
\approx &  
e^{- \frac{N I^2}{6} \int da p(a) \left( a {\bf x} + {\bf y} \right)^2}
=
e^{- \frac{N I^2}{6} \left( \overline{a^2} {\bf x}^2 + 2 {\bar a} {\bf x} \cdot {\bf y} + {\bf y}^2 \right)}
\; .
\end{split}
\end{equation}
Finally we integrate over the Lagrange multipliers ${\bf x,y}$, and get
\begin{equation}
{\mathcal Z}_1 =
\int d{\hat{\bf m}} 
e^{
- \frac{3}{2} \frac{
\left(
 {\sf H}_N^2 + \overline{a^2} {\boldsymbol{\sf I}}^2 - 2 \bar{a}
{\boldsymbol{\sf H}_N}  \cdot \boldsymbol{\sf I}
\right)
}{N I^2(\overline{a^2}-\overline{a}^2)}}
\; .
\end{equation}
We define the probability distribution as follows
\begin{equation}
{\mathcal Z}_1 =
\int d\cos(\theta) P_1(\cos{\theta})
\; ,
\end{equation}
where $\theta$ is the angle between ${\hat m}$ and $\hat{\boldsymbol{\sf I}}$.
The distribution function then takes the form
\begin{equation}
P_1(\cos{\theta}) = e^{\alpha \cos{\theta}}
\; ,
\end{equation}
where $\alpha = \frac{3 \bar{a}{\sf H}_N {\sf I}}{NI^2(\overline{a^2}-\bar{a}^2)}$ 
as defined in Sec.~\ref{sec:appr-treatm-time}.

Similarly, simply introducing another Lagrange multiplier, we can find
\begin{equation}
P_2(\cos{\theta}) = e^{\gamma \cos{\theta} + \beta {\cos^2{\theta}}}
\;,
\end{equation}
where $\beta$ and $\gamma$ are defined in Sec.~\ref{sec:appr-treatm-time}.

For $P_p(\cos{\theta})$ when $p>2$, the calculation becomes more and more
complicated, and hard to get an analytical expression.

\end{document}